\documentclass[useAMS,usenatbib]{mn2e}

\usepackage{graphicx}
\usepackage{hyperref}
\usepackage{booktabs}

\title[Star Formation and Clumps in Radiation Pressure Simulations ]{Star Formation and Clumps in Cosmological Galaxy Simulations with Radiation Pressure Feedback}
\author[Christopher E. Moody]{Christopher E. Moody$^{1}$, Yicheng Guo$^{2}$, Nir Mandelker$^{3}$, Daniel Ceverino$^{4}$,
\newauthor Mark Mozena$^{2}$, David C. Koo$^{2}$,  Avishai Dekel$^{3}$, and 
Joel Primack$^{1}$\thanks{E-mail: joel@ucsc.edu} \\
$^{1}$Department of Physics, University of California, Santa Cruz, CA, USA \\
$^{2}$UCO/Lick Observatory, Department of Astronomy and Astrophysics, University of California, Santa Cruz, CA 95064, USA \\
$^{3}$Racah Institute of Physics, The Hebrew University, Jerusalem 91904 Israel\\
$^{4}$Departamento de F\'isica Te\'orica, Universidad Aut\'onoma de Madrid, 28049 Madrid, Spain}

\begin{document}
\addtolength{\topmargin}{-0.7in}

\date{5 May 2014}


\maketitle

\label{firstpage}

\begin{abstract}
Cosmological simulations of galaxies have typically produced too many stars at early times. We study the global and morphological effects of radiation pressure (RP) in eight pairs of high-resolution cosmological galaxy formation simulations. We find that the additional feedback suppresses star formation globally by a factor of $\sim2$. Despite this reduction, the simulations still overproduce stars by a factor of $\sim2$ with respect to the predictions provided by abundance matching methods for halos more massive than $5\times10^{11}M_{\odot}h^{-1}$ \citep{2013ApJ...770...57B}. 

We also study the morphological impact of radiation pressure on our simulations. In simulations with RP the average number of low mass clumps falls dramatically.  Only clumps with stellar masses $M_{clump}/M_{disk}\leq5\%$ are impacted by the inclusion of RP, and RP and no-RP clump counts above this range are comparable. The inclusion of RP depresses the contrast ratios of clumps by factors of a few for clump masses less than 5\% of the disk masses. For more massive clumps, the differences between and RP and no-RP simulations diminish. We note however, that the simulations analysed have disk stellar masses below about $2 \times 10^{10} M_\odot h^{-1}$. 

By creating mock Hubble Space Telescope observations we find that the number of clumps is slightly reduced in simulations with RP.  However, since massive clumps survive the inclusion of RP and are found in our mock observations, we do not find a disagreement between simulations of our clumpy galaxies and observations of clumpy galaxies. We demonstrate that clumps found in any single gas, stellar, or mock observation image are not necessarily clumps found in another map, and that there are few clumps common to multiple maps.  

\end{abstract}


\section{Introduction}

Star-forming galaxies in the redshift range $z\sim1-3$ are frequently observed with thick, turbulent disks and giant clumps. With the SINS survey the morphology of high-{\it z} galaxies has drawn considerable attention and revealed that high-mass galaxies show ordered rotation despite hosting these large clumps \citep{2011ApJ...733..101G,2009ApJ...706.1364F}. 
Furthermore, the clumps also appear to be morphologically significant, typically being $\sim$kpc in size, and emitting half the rest-frame ultraviolet light \citep{Elmegreen:2005ub,ForsterSchreiber:2006cy, Genzel:2008dt}. Observations using the Hubble Space Telescope have resolved sub-kiloparsec scales and measured the detailed properties of clumps, showing that 30\% of SFR of these galaxies is in the form of clumps, with individual clumps contributing $\sim10\%$ of the total SFR at $z\sim2$ \citep{2012ApJ...753..114W,2012ApJ...757..120G,2012ApJ...749..149G}. With masses of $\sim10^7-10^9M_\odot$, these clumps are much larger than local star-forming molecular clouds that have masses of $\sim10^5-10^6M_\odot$. The clumpy morphology of high-redshift galaxies is thus markedly different from local galaxies, and has precipitated further studies.  

Attempting to match observations, recent theoretical analyses have targeted this early epoch of galaxy formation. In both isolated and cosmological simulations, clumps form from gravitational instabilities within a gas-rich turbulent disk without associated dark matter halos \citep{2007ApJ...670..237B,2010MNRAS.404.2151C}. In detailed studies of multiple simulated galaxy histories the fraction of clumpy disks peaks at $z\sim2$, with 1\%-7\% of the disc mass in the form of clumps but comprising 5\%-45\% of the star formation rate \citep{2013arXiv1311.0013M}, demonstrating broad consistency with observations. Simple theoretical frameworks argue that these clumps, while individually short-lived, are formed in a steady-sate disk that is continually replenished by cold streams \citep{2013MNRAS.435..999D,2009ApJ...703..785D}. Within a few orbital times, the clumps coalesce onto the bulge.   

Despite successes in reproducing observed clump properties, cosmological simulations generically find that stars are over-produced \citep[e.g.,][]{2012MNRAS.426.2797W} with respect to new constraints on the stellar mass-halo mass relationship \citep{2010ApJ...717..379B,2013ApJ...770...57B,2010ApJ...710..903M,2013MNRAS.428.3121M}. These constraints assume that every dark matter halo or subhalo above a mass threshold hosts one galaxy. These results suggest that the peak of the stellar mass-halo mass relation occurs around $10^{12}M_\odot$ with $\sim2\%$ of the total mass in stars, falling to $\sim0.4\%$ for less massive $10^{11}M_\odot$ halos, and $0.6\%$ for more massive $10^{13}M_\odot$ halos. To match these constraints, galaxy simulations have adopted new forms of stellar feedback to suppress or stop star formation. In particular, recent simulations with radiation pressure feedback have had success in depressing the star formation rate at early times and especially in low-mass halos \citep{Ceverino:2013tt, 2012MNRAS.421.3522H,2013ApJ...770...25A,2011ApJ...735...66M,Hopkins:2013ua}. However, the additional injection of pressure in small scales opens the possibility of changing the morphology at the $\sim$kpc scale, and possibly disrupting or preventing the clump formation altogether. 

This paper tests whether radiation pressure feedback maintains an appropriate stellar mass to halo mass ratio while still sufficiently preserving clumps to match abundance models. The outline of this paper is as follows. In Section 2 we describe the simulations and analysis methods. This discussion includes description of the galaxy simulations with and without radiation pressure feedback as well as the details of the dust modelling in the radiative-transfer simulations. We also describe the analysis methods, halo and sub-halo finding techniques, and clump finding techniques. In Section 3 we detail the global effects of radiation pressure on simulated galaxies, particularly focusing on the relationship between stellar mass and halo mass at various epochs. Having discussed the global effects of radiation pressure, in Section 4 we focus on the changes in clump morphology, specifically on clump mass and number statistics. In Section 5 we present results from mock observations of the simulations, thereby translating from physical quantities accessible only in simulations to directly observable quantities. In Section 6 we summarise our conclusions.

\section{Analyzing the Simulations}

\subsection{The ART Simulations}
Our sample consists of eight pairs of galaxies simulated using the Adaptive Mesh Refinement (\texttt{ART}) code \citep{Kravtsov:1997p3241,2003ApJ...590L...1K,2009ApJ...695..292C}. The simulation code incorporates many of the physical processes relevant for galaxy formation, including gravitational N-body dynamics, Eulerian hydrodynamics, photoionisation heating, star formation, stellar mass loss, stellar feedback, and metal enrichment as described in \citet{2009ASPC..419..410C,2010MNRAS.404.2151C}.  The cooling rates are computed for a given gas density, temperature, and metallicity and include the effect of UV attenuation due to gas self-shielding at high densities. No active galactic nucleus (AGN) feedback is included, although for our relatively low halo mass range ($10^{11} - 10^{12} M_\odot$) this is not likely to be a dominant effect. The simulations feature a high dark matter mass resolution of $8\times10^4M_{\odot}$ and an adaptive mesh refinement resolution of 17-35pc, which is sufficient to resolve typical small stellar clusters.

\begin{table}

\begin{tabular}{llrr}
\toprule
Simulation & $M_{halo}\,(M_{\odot}h^{-1})$ &  $M_*\,(M_{\odot}h^{-1})$ &  $M_{disk}\,(M_{\odot}h^{-1})$ \\
\midrule

 VELA02RP &         $1.99 \times 10^{11}$ &      $2.33 \times 10^{9}$ &           $0.69 \times 10^{9}$ \\
    VELA02 &         $1.96 \times 10^{11}$ &      $4.44\times 10^{9}$ &           $0.70 \times 10^{9}$ \\
 VELA03RP &         $2.13 \times 10^{11}$ &      $5.10 \times 10^{9}$ &           $0.67 \times 10^{9}$ \\
    VELA03 &         $2.17 \times 10^{11}$ &     $8.57\times 10^{9}$ &           $1.91 \times 10^{9}$ \\
 VELA05RP &         $1.13 \times 10^{11}$ &      $1.26 \times 10^{9}$ &           $0.31 \times 10^{9}$ \\
    VELA05 &         $1.11 \times 10^{11}$ &      $3.44 \times 10^{9}$ &           $1.11 \times 10^{9}$ \\
  VELA13RP &         $4.55 \times 10^{11}$ &     $12.26 \times 10^{9}$ &           $6.96 \times 10^{9}$ \\
    VELA13 &         $4.52 \times 10^{11}$ &     $19.11 \times 10^{9}$ &           $13.80 \times 10^{9}$ \\    
 VELA14RP &         $5.00 \times 10^{11}$ &     $17.44 \times 10^{9}$ &           $3.57 \times 10^{9}$ \\
    VELA14 &         $4.76 \times 10^{11}$ &     $17.71 \times 10^{9}$ &           $5.83 \times 10^{9}$ \\
 VELA26RP &         $5.24 \times 10^{11}$ &     $18.24\times 10^{9}$ &           $6.23 \times 10^{9}$ \\
    VELA26 &         $5.15 \times 10^{11}$ &     $24.64 \times 10^{9}$ &          $15.14 \times 10^{9}$ \\
 VELA27RP &         $4.42 \times 10^{11}$ &     $10.04 \times 10^{9}$ &           $3.39 \times 10^{9}$ \\
    VELA27 &         $4.39 \times 10^{11}$ &     $18.11\times 10^{9}$ &           $10.43 \times 10^{9}$ \\
 VELA28RP &         $2.76 \times 10^{11}$ &      $3.13 \times 10^{9}$ &           $0.57 \times 10^{9}$ \\
    VELA28 &         $2.78 \times 10^{11}$ &     $8.13 \times 10^{9}$ &           $1.09 \times 10^{9}$ \\
    
\bottomrule
\end{tabular}

\caption{For each simulation the name is given as well as the halo virial mass at redshift $z=2$ and the stellar mass inside a tenth of the virial radius at $z=2.$}
\label{table1}
\end{table}

Halos are selected in the virial mass range of $10^{11} - 10^{12} M_\odot$ and to have no ongoing major merger at $z=1$. The latter criterion removes 10\% of halos but otherwise has no obvious ramifications for the formation of history of halos at $z>2$. Having randomly selected a halo within the desired mass range and targeted a zoom-in region, the simulations are rerun with full physics enabled inside the high-resolution region.  All galaxies are then evolved to a redshift of $z\sim1$. 

A list of the simulation names, central halo masses,  stellar masses and stellar disk masses at z=2 are given in Table \ref{table1}. We use the \texttt{ROCKSTAR} halo finder \citep{2013ApJ...762..109B} to calculate both the virial radius and mass. For the halo masses, \texttt{ROCKSTAR} calculates the spherical overdensity using the virial density threshold from \citet{Bryan:1998te}. This calculation includes all of the substructures in a halo. Also given is the stellar mass within a tenth of the virial radius of each halo.  The disk mass is the total mass of stars with at least 70\% of the angular momentum required for a circular orbit. The details of the disk mass calculation are discussed in Section \ref{disc_mass}.

The simulation snapshots are available in scale factor increments of $0.01$ which separates consecutive snapshots by approximately 120Myr, although this spacing grows more dense at later times. This time spacing is of order the typical orbital time and so we assume each snapshot is an independent sample. For all of the relevant analyses, snapshots are organised into pairs of radiation pressure (RP) and no radiation pressure (no-RP) snapshots. Snapshots are analysed pairs at a time. In this way, extrinsic quantities such as clump counts are comparable and well-balanced. If data is missing such that a single snapshot of a pair is not available, the whole pair is removed from subsequent time steps in order to maintain a fair data sample. 

\subsection{The Two Feedback Models}
The purpose of this paper is to investigate the global and morphological effects of radiation pressure feedback in galaxies around $z\sim2$. We test two models of stellar feedback, the first (no-RP) is a purely thermal mode of feedback arising from supernovae and stellar winds. This base feedback expresses the energy released by supernova as a constant heating rate spread out over 40 Myr following a star formation event. The second feedback model introduces additional non-thermal pressure representative of the radiation pressure around dense and optically thick star-forming regions. Only the ionising radiation, instead of the full stellar luminosity, is considered, thereby restricting the effect of RP to the first $\sim5$Myr of the life of a single stellar population. The momentum imparted by the radiation is locally injected as additional pressure in a single cell, neglecting radiation transfer effects. The large optical depths in these regions can trap UV radiation, limit the gas supply available to form stars, and thus lead to a self-regulated star formation rate \citep{Ceverino:2013tt}. The global effect of this additional feedback is to curtail star formation, especially at high redshifts, leading to stellar-mass-to-halo-mass ratios closer to abundance matching estimates. For implementation details and the underlying models refer to \citet{Ceverino:2013tt}.  Other simulations show that our RP implementation also produces realistic star formation histories for lower mass galaxies \citep{2013arXiv1311.2910T}.

\subsection{Analysis Methodology and Clump Finding}
\begin{figure*}
\includegraphics[scale=1.00]{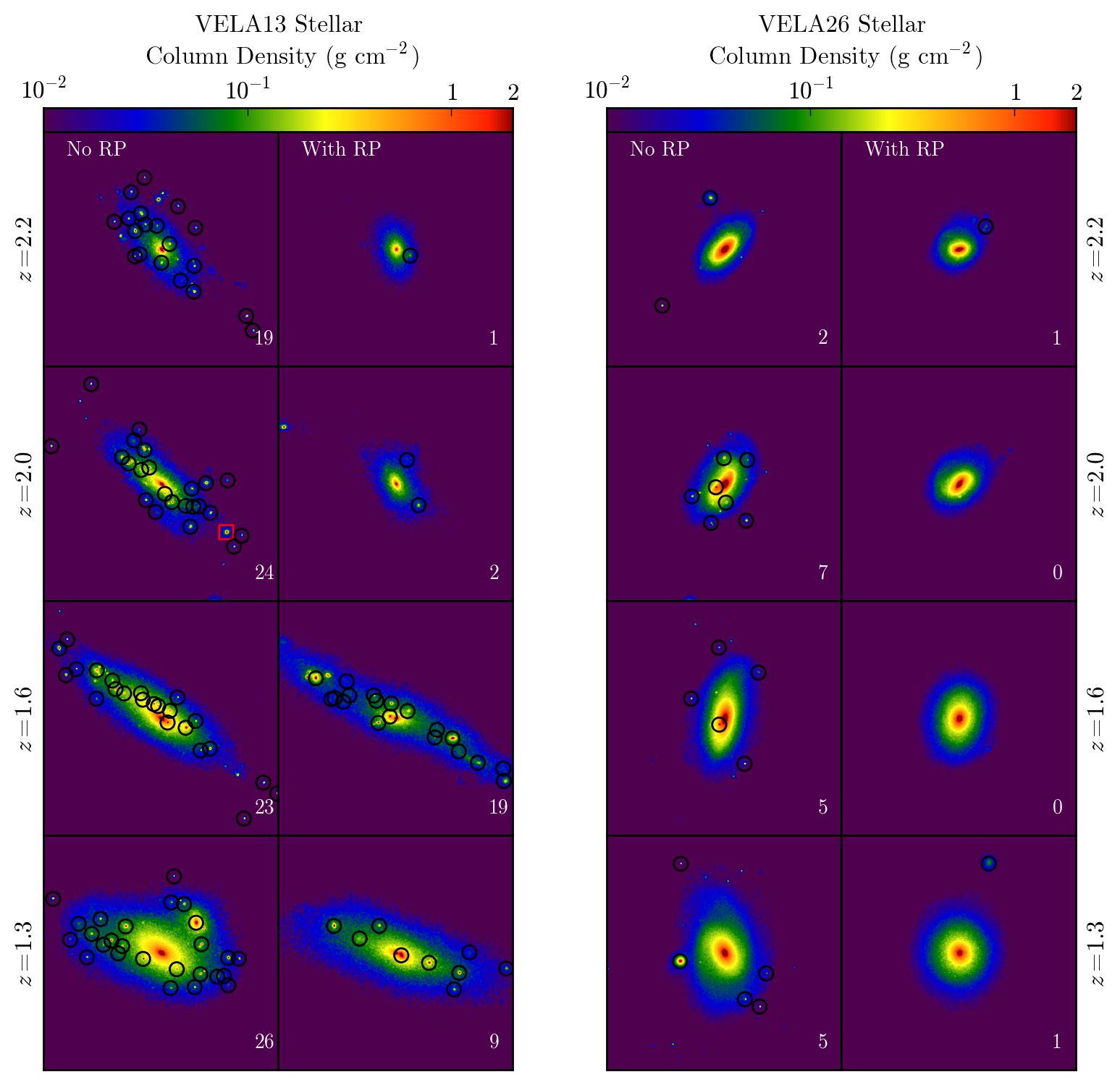}
\caption{The projected stellar mass densities of two pairs of RP and no-RP simulations are plotted above at a series of redshifts between $z=1.3$ and $z=3.0$. Detected clumps (black circles) and minor mergers are marked (red squares) and the total number of clumps is inlaid (white text). In the first and second columns the no-RP and RP versions of  the `VELA13' simulation are plotted. In the second and fourth columns the no-RP and RP versions of the `VELA26' simulation are plotted.}
\label{fig:epsm}
\end{figure*}
In this section we discuss the analysis methods concerning data presented in Sections \ref{sec:gp} and \ref{sec:cd}. Our analysis pipeline consists of first finding halos, isolating a main progenitor lineage in the galaxy's formation history, preparing projected stellar mass maps, and finally finding clumps within these projected maps. We use the \texttt{ROCKSTAR} halo finder to identify dark matter halos using just the highest resolution, lowest mass, dark matter particles \citep{2013ApJ...762..109B}. \texttt{ROCKSTAR} searches for clusters of particles by identifying neighbours in six phase-space dimensions via friends-of-friends. For every group the linking distance is adaptively renormalised across the six dimensions, yielding a parameter-free and natural distance calculation. By investigating the phase space kinematics \texttt{ROCKSTAR} has the ability to distinguish a pair of cospatial mergers in very dense environments. This is especially crucial in discriminating clumps formed in the disk from minor mergers, where the high dark matter densities at the centre of the host halo may otherwise obscure an infalling satellite. 

Once the halos and subhalos are identified and their basic properties are measured, we iteratively find the most massive progenitor history while trying to preserve halo properties across timesteps using the publicly available \texttt{consistent-trees} software \citep{2013ApJ...763...18B}. In addition to applying this consistency check we also normalise the halo merger trees by demanding that halos do not suddenly move unphysically large distances ($>3\%$ of the box size) in a single timestep. In such cases the halo finder has likely misidentified a halo from its particle membership, and we reassign the progenitor to be the nearest halo with reasonably consistent properties. Once all the progenitor halos are identified, we find the centre of each halo by computing peak stellar density within the scale radius of the halo.

Two-dimensional maps of the projected stellar mass, each $20$kpc/h wide and deep, are then computed for each halo. The images are binned onto 600 pixels, giving a width of 32pc per pixel. For computational convenience, the maps are projected along the code axes instead of face-on or edge-on axes. We use the \texttt{yt} library to load in the octree hierarchy in each \texttt{ART} snapshot \citep[][and see \url{http://yt-project.org}]{2011ApJS..192....9T}.  For every particle in the snapshot we deposit the relevant quantities (e.g. mass) into each cell of the octree, effectively transforming discrete Lagrangian variables onto a spatially continuous Eulerian mesh. As it is a computationally expensive process, we do not consider the SPH kernel of each particle and instead deposit all of the particle properties locally and directly into the nearest cell instead of smoothing the properties over a local neighbourhood of cells. As a result, our three-dimensional local density estimate can suffer from increased Poisson noise, yielding a poorer signal-to-noise ratio than if we had implemented a smoothing kernel. We note that since we are restricting ourselves to two-dimensional projections instead of analysing the three-dimensional octree, the effect of noise is mitigated.  Additionally, by restricting ourselves to two dimensional images our analysis becomes more comparable to observational analyses. Having constructed the mass maps, we then prepare them by smoothing them with a Weiner filter, which attempts to denoise an image. This has the effect of further suppressing noise fluctuations below the scale of a few pixels. In practice, and unlike Gaussian smoothing, we find that the Weiner filter better preserves the peak densities while still smoothing regions with large stochastic variation. An example of resulting images is shown in Figure \ref{fig:epsm}.

\label{clumpfinder}
We then contour the image at a set of thresholds. We identify any contour that includes the centre of the host galaxy, and remove it from further analysis. Rarely, noise patterns can conspire to create a falsely-identified clump. As a result, we filter out the smallest clumps by enforcing that the clump surface area must be greater $0.02$ kpc$^2$ corresponding to a sphericalised radius of 80pc. We also remove spurious and non-spherical clumps by enforcing the shape criterion that a clump must contain its own centre of mass.  In practice the rejected contour regions enclose very little mass and excluding them as false positives removes fewer than $3\%$ of clumps. Despite the heuristic filters described so far, the clump finding algorithm can still fail in certain high-noise environments and can occasionally find an excess of clumps. Furthermore, low peak density clumps below a peak threshold of $3\times10^{-2}$g cm$^{-2}$ are not detected. This limit corresponds to approximately $\sim100$ stellar particles per pixel. As a result, we cannot resolve clumps below a mass limit of $\sim10^6M_{\odot}$. After the stellar clumps are found we calculate the nearest halo, including subhalos, and tag the clump as being ex-situ (e.g., a minor merger) if a subhalo centre is less than $100pc$ from the clump centre.

Clump properties are then calculated as a function of cells enclosed in each contour. Once the region enclosing each clump is found, the mass of that clump is calculated as the projected stellar mass of each enclosed pixel. This may overestimate the clump mass and other properties because the underlying background disk mass is included. To mitigate this effect we can estimate and subtract off the disk contribution. We have experimented with controlling for the effect of a background by identifying the outskirts of the clump region and extrapolating its effect to the full clump. In a measurement of the mass, for example, this would constitute a measurement of the circumferential density, multiplying by the clump surface area to estimate the background disk mass, and subtracting this sum from the total region mass. In these experiments we find that clump mass is typically reduced by $\sim1-2\%$.  This reinforces the notion that the clump masses are highly centralized and that the disk does not strongly contribute to the clump mass over the small extent of the clump. As a result of these experiments, we do not attempt to control for the background effect of the disk in any way.

\label{disc_mass}
We use the \texttt{yt} code \citep{2011ApJS..192....9T} to calculate the stellar disk mass, angular momentum vector, and the stellar mass enclosed in a tenth of the virial radius. This radius is chosen both to avoid including in-falling satellites and to match the stellar mass definition found in the \citet{2013ApJ...770...57B} abundance matching models. The disk mass is calculated as the sum of stars with angular momentum exceeding $70\%$ of the angular momentum required to maintain a circular orbit at that radius, $\epsilon = j_{z} / j_c > 0.7$. This criterion is identical to that used in other recent papers \citep{2011ApJ...742...76G,2013ApJ...772...36G,2010MNRAS.404.2151C}. Here, the $j_{z}$ is the angular momentum of the star particle in the direction of the galactic angular momentum vector,  and $j_c(r)$ is the orbital angular momentum, $j_c(r)=m r V_c(r)$, computed using the circular velocity, $V_c(r)=(GM(<r)/r)^{\frac{1}{2}}$, at a given radius. The star particle mass is denoted by $m$, the radius by $r$, and $M(<r)$ denotes the total baryonic and dark matter mass enclosed by the particle.

The code and scripts used to create the figures in this paper are publicly available at \url{http://bitbucket.org/juxtaposicion/rpa}.

\subsection{Sunrise Mock Images and Clump Finding}
In this section we discuss the analysis methods concerning data presented in Section \ref{sec:mock}. We use the \texttt{SUNRISE} code to generate realistic images directly comparable to observed Hubble Space Telescope (HST) images \citep{2006MNRAS.372....2J,2010MNRAS.403...17J,2010NewA...15..509J}. The code performs a Monte Carlo radiation transfer calculation through the stellar geometries supplied by the \texttt{ART} snapshots. We assume 40\% of the gas-phase metals are in the form of dust. The dust grain model used is the R = 3.1 \cite{2001ApJ...548..296W} dust model including updates by \cite{Draine:2007p1267}. We use the \texttt{SUNRISE} interface provided by the \texttt{yt} code to facilitate the conversion process. Importantly, the effect of the dusty ISM is included in the \texttt{SUNRISE} calculation as photon packets are absorbed and scattered by intervening material. Young stellar particles are assumed to represent HII and photodissociation regions around starforming clusters and their spectra are computed using  MAPPINGS models \citep{2008ApJS..176..438G}. The resulting image realistically reproduces the spectral energy distribution from the ultraviolet to the submillimetre in each pixel, allowing us to test galaxy formation simulations directly with observed images. While \texttt{SUNRISE} can self-consistently calculate the dust temperature and thus model far-IR emission, we do not attempt to analyse these wavelengths and instead restrict ourselves to studying mock images in  the UV-optical bands selected for use in the CANDELS survey \citep{2011ApJS..197...36K,2011ApJS..197...35G}. Once the images have been calculated we add a noise background, repixelise, and blur the image using a point-spread function (PSF) in order to recreate the noise properties typical of CANDELS observations (M. Mozena in preparation and P. Kollipara in preparation). The noise backgrounds, pixelisation and point-spread function are designed to match the properties of the Hubble Space Telescope V, i, z and H filters. We refer to images produced by this process as `V-band Mock', `i-band Mock' , `z-band Mock', and `H-band Mock'.

\begin{figure*}
\includegraphics[scale=0.90]{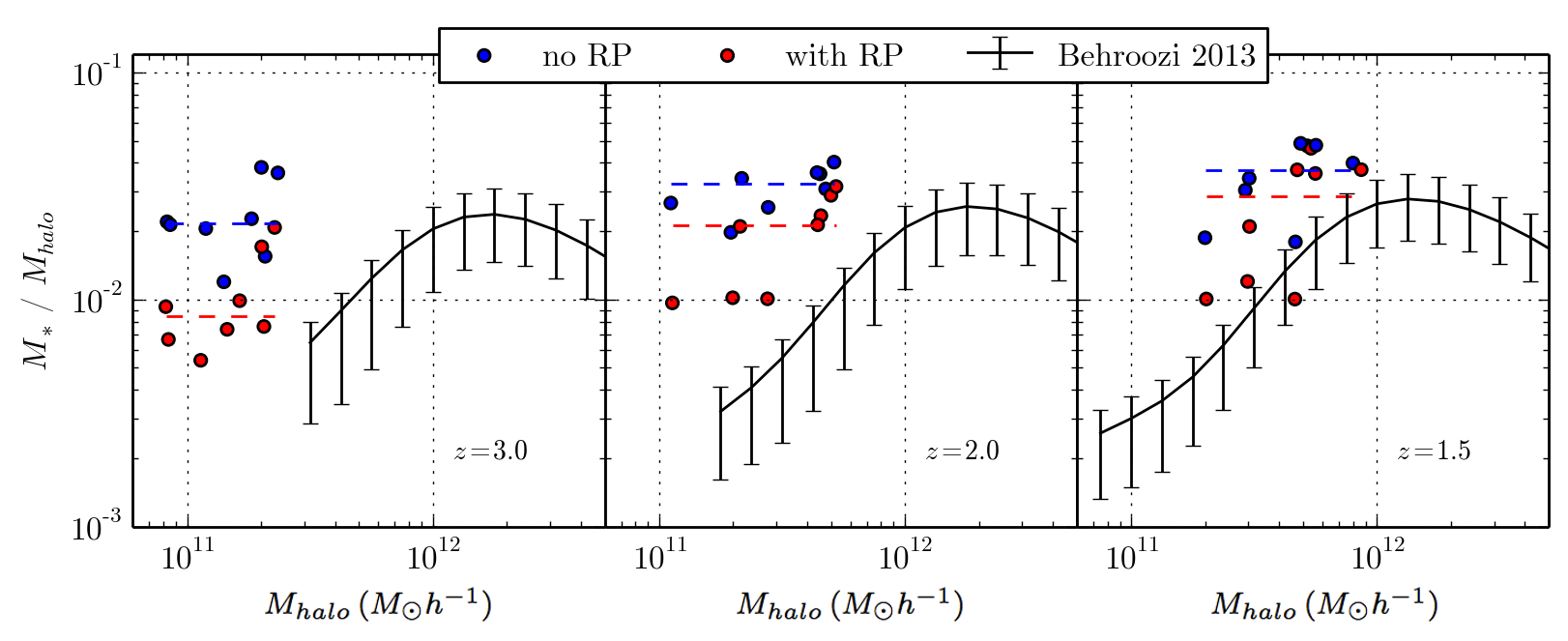}
\caption{The stellar mass to halo mass ratio as a function of halo mass is plotted for radiation pressure (RP, red) and no radiation pressure (no RP, blue) simulations at redshifts of $z=3$ (left), $z=2$ (middle), and $z=1.5$ (right).  At each redshift, the median ratio is shown for RP (dashed red lines) and no-RP simulation (dashed blue lines). RP simulations depress the stellar mass by a factor of  $\sim2-3$ at all redshifts.  For comparison the stellar mass function and 32-68$\%$ percentile uncertainty about the median is shown from \citet{2013ApJ...770...57B}. }
\label{fig:msmh}
\end{figure*}

\section[]{Global Properties}
\label{sec:gp}
In Figure \ref{fig:msmh} the $M_{star}-M_{halo}$ relation for RP simulations, no-RP simulations, and abundance matching predictions \citep{2013ApJ...770...57B} are shown at three different redshifts, $z=3, 2, 1.5$ corresponding to times when the universe was $2.2, 3.3, 4.3$ Gyr old, respectively. For the predicted abundance matching data, the central data point denotes  the median stellar mass ratio, while the error bars give a $1\sigma$ confidence interval. The primary effect of RP is to suppress the mass of formed stars by a factor of $\sim2$ at all redshifts, but especially so at early times. 


We define the overproduction of stars as the simulated $M_{star}/M_{halo}$ divided by the $M_{star}/M_{halo}$ from abundance matching models.
The abundance matching predictions at $z=3$  (left panel) do not extend to our low halo mass range.  
We instead use the closest available data point at $M_{halo}=3\times10^{11}M_\odot, M_{star}/M_{halo}=0.6\%$ as the reference point.  This is equivalent to assuming that the $M_{star}/M_{halo}$ is flat at halo masses $M_{halo}<3\times10^{11}M_\odot$ for $z=3$.
For each snapshot at this redshift we compare the simulated $M_{star}/M_{halo}$  to that given by the abundance matching models, finding that the median no-RP (RP) simulation overproduces stars by a factor 3.6 (1.6).  However, if the stellar abundance ratios continue to decline at lower halo masses, this discrepancy will be underestimated.  
No-RP simulations in this redshift range have a median $M_{star}/M_{halo}$ of  2.2\%,  while the median RP simulation has $M_{star}/M_{halo}=0.84\%$. 



At redshift of $z=2$  (middle panel) the abundance matching data extend to low enough masses to provide direct comparisons for all but one of our simulations. 
The median no-RP (RP) simulation has $3.2\%$ (2.1\%) of the halo mass in the form of stars, and overproduces stellar mass by a factor of 5.5 (3.1).  
At redshift $z=1.5$ (right panel), the median no-RP (RP) simulation has $M_{star}/M_{halo}=3.7\%$ ($M_{star}/M_{halo}=2.8\%$)  and overproduces stars by a factor of 2.8 (1.9).
The addition of RP feedback on depresses the median stellar mass by 2.4, 1.8, and 1.6 when compared to no-RP simulations at redshifts of 3, 2, and 1.5, respectively. The effect of RP is stronger at early times and diminishes with time. 

While our simulations begin to show an upturn in the $M_{star} - M_{halo}$ relation, we do not attempt to simulate halos beyond the peak halo mass ($\sim2\times10^{12} M_\odot$). Semi-analytic models and simulations suggest a need to quench the star formation rate of halos with masses greater than this peak mass with `radio-mode' AGN feedback \citep{2006MNRAS.365...11C}.  Because we do not include any such quenching mechanism we do not explore simulations in this mass range. Despite the additional feedback, the RP simulations still overproduce stars and this fact suggests that increasing the feedback will bring closer agreement with stellar mass abundance matching.

\section[]{Clump Distributions}
\label{sec:cd}
In this section we discuss the distribution of clump properties in simulated 2D mass maps and characterise the effect of RP with respect to clumps.  As we discuss in this section, the dominant morphological effect of RP is to reduce the number of low-mass clumps. However, realistic simulations must reproduce the observed prevalence of clumpy galaxies at high redshift. In this section we report on the distribution of contrast ratios, and the distribution of average number of clumps. In each of these cases we contrast these quantities with respect to clump mass since that is the that is the directly computed quantity. However, some theoretical models expect that the clump growth is mediated by the disk mass and so we plot the clump mass normalised by the host disk mass. Finally, we also plot the clump mass normalised by the total stellar mass as this quantity is more easily inferred in observations than the disk mass. Normalising to the disk mass and total stellar mass allows us to control for the effect of RP. We conclude that the effect of RP is chiefly on low-mass clumps with $M_{clump} / M_{disk} \leq 5\%$ and that the properties of no-RP and RP clumps converge for high mass clumps.

\begin{figure*}
%
\includegraphics[scale=1.0]{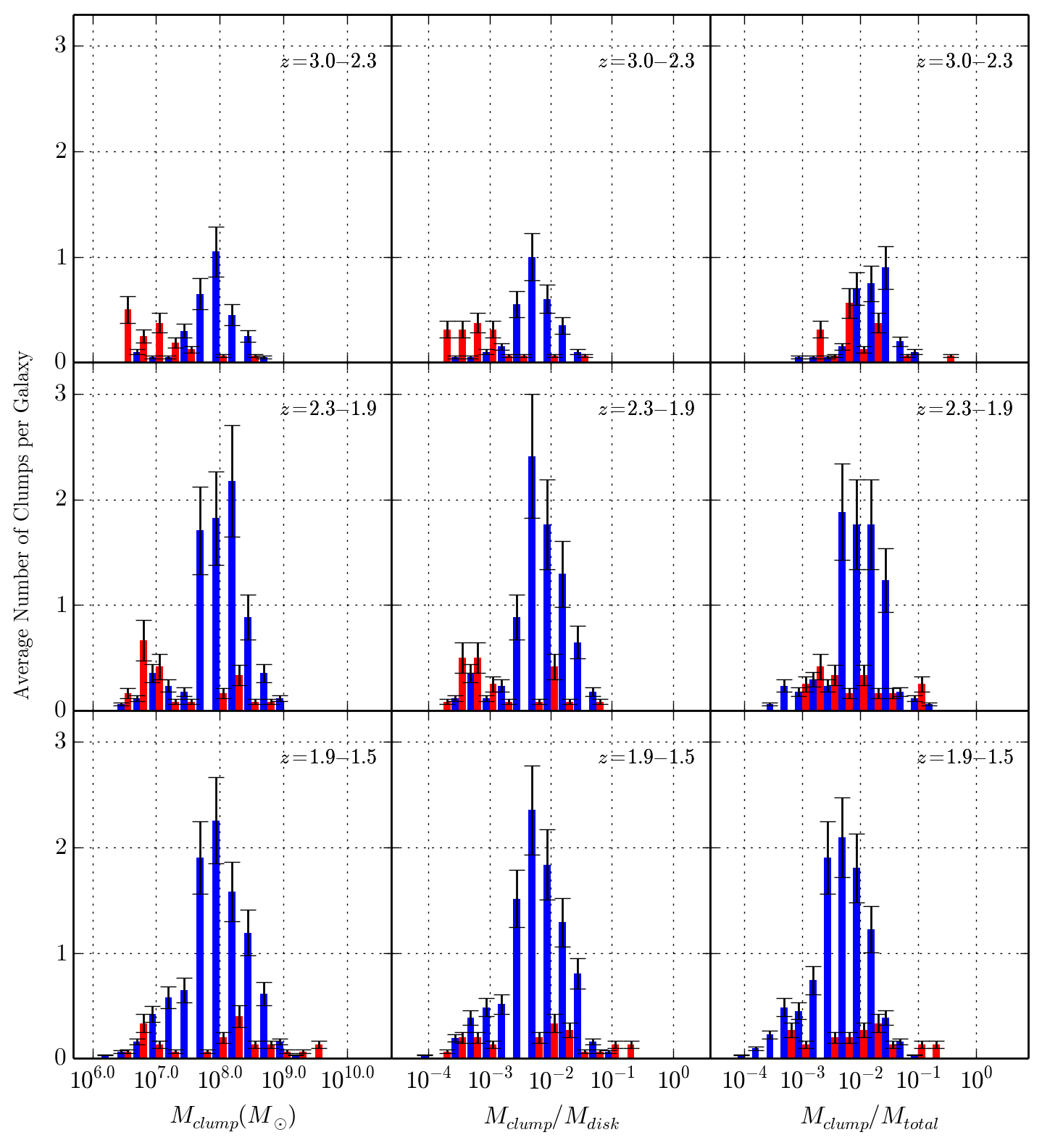}
\caption{The average number of clumps is shown as a function of the clump stellar mass (left column), clump stellar mass normalised to the stellar mass of the disk in the host galaxy (middle column), and the clump stellar mass normalised to the total stellar mass of the host galaxy (right column). Simulations are divided into RP (red) and no-RP (blue) classes and shown for three redshift ranges, $z=3.0-2.3$ (first row), $z=2.3-1.9$ (second row), and $z=1.9-1.5$ (third row).  In most mass bins, regardless of which normalisation is used, no-RP simulations produce more clumps than RP simulations.  The ratio between the average number of no-RP and RP clumps peaks for intermediate-mass clumps. This ratio becomes less dramatic for clump masses larger than $M_{clump}/M_{disk}>5\%$. Overall, the average number of clumps per galaxy grows in time with the redshift interval $z=3.0-2.3$ hosting fewer clumps than later redshifts.}
\label{fig:ncmcmd}
\end{figure*}

In Figure \ref{fig:ncmcmd} we plot the distribution of the average number of clumps per galaxy for a given clump mass for three redshifts  $z=3.0-2.3$ (first row), $z=2.3-1.9$ (second row), and $z=1.9-1.5$ (third row). Three definitions of the clump mass are used. The clump mass $M_{clump}$ (left column) is the stellar mass content of each clump, $M_{clump}/M_{disk}$ (middle column) is the clump mass normalised to the disk mass, and $M_{clump}/M_{total}$ (right column) is the clump mass normalised to the total galaxy stellar mass. As usual, RP simulation snapshots (red) are always paired with no-RP simulation snapshots (blue) to ensure a fair comparison. 

The average number of clumps per galaxy in the simulations grows as time goes on. The average number of clumps of all masses per galaxy for no-RP simulations climbs from 2.9 at the highest redshift range,  $z=3.0-2.3$, to 8.0 and 9.6 for redshift ranges $z=2.3-1.9$ and $z=1.9-1.5$ respectively.  RP simulations at all times host fewer clumps, starting with 1.6 average clumps per galaxy over all masses at $z=3.0-2.3$. At later times, RP simulations host 2.1 and 1.8 clumps on average per galaxy over all masses for redshift ranges $z=2.3-1.9$, and $z=1.9-1.5$ respectively. The average number of clumps in no-RP simulations grows quickly, while the growth in number of clumps in RP simulations is weaker and slower.
 

The three columns of Figure \ref{fig:ncmcmd} demonstrate the breakdown of these clumps into different stellar mass bins. These three measures correspond to the direct simulation clump mass, $M_{clump}$, the theoretically motivated driver for clump growth, $M_{clump}/M_{disk}$, and the clump quantity most accessible in observations,  $M_{clump}/M_{total}$. The distributions of clump masses in no-RP simulations are frequently peaked in a single mass bin. In contrast, no-RP distributions more uniformly span the range of masses, with no obvious peaks in the clump mass distribution.  The peak $M_{clump}$ at all times occurs in the $10^8M_{\odot}$ mass bin. The peak normalised clump masses $M_{clump}/M_{disk}$ also stays fixed in time and in the $M_{clump}/M_{disk}=10^{-2.25}$ bin. Over time, the clump mass normalised by the total galaxy stellar content shifts to lower values, peaking $M_{clump}/M_{total}=10^{-1.5}$ at early times but shifting to $M_{clump}/M_{total}=10^{-2.25}$ for later redshift ranges.


We define the RP to no-RP ratio as the average number of RP clumps divided by the number of no-RP clumps in a given mass bin. This ratio then captures the relative excess of no-RP clumps compared to RP clumps. At low masses, $M_{clump}/M_{disk} < 10^{-3}$, the average number of clumps is small and comparable between RP and no-RP cases. Therefore, this ratio is close to unity for very low-mass clumps. For clump masses near the peak no-RP clump mass, the number of RP clumps stays approximately constant while the number of no-RP clumps climbs. For these masses, the RP to no-RP difference is the largest.  Around high masses, $M_{clump}/M_{disk} > 10^{-1.25} \sim 5\%$, this ratio reaches unity again. Since the average number of clumps in no-RP simulations equals that in RP simulations, we speculate that around this mass range the clump self-gravity balances the outward radiation pressure to stabilise the clump. For clumps smaller than this mass range we only know that instantaneous clump counts are not equal. We do not follow the history of individual clumps but instead focus on the properties of clumps available instantaneously. As a result, we cannot distinguish between scenarios where radiation pressure prevents a clump from ever forming, or where a clump forms and subsequently dissipates on short timescales. 



\begin{figure*}
\includegraphics[scale=0.85]{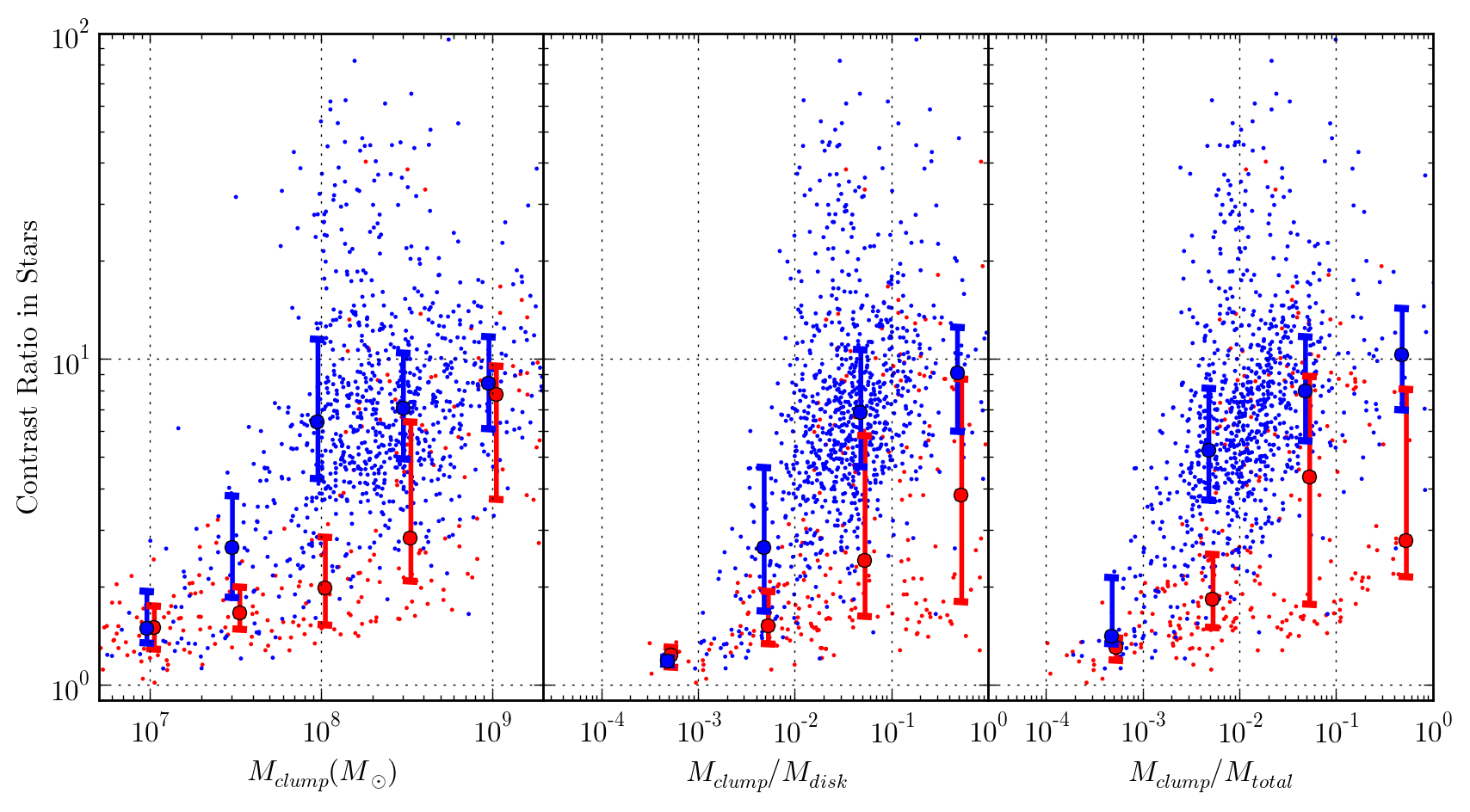}
\caption{The contrast ratio of every detected clump in our sample versus the mass of that clump is shown. Error bars are drawn denoting the $25\%$, $50\%$ and $75\%$ percentiles of the contrast ratio within a clump mass bin. Each clump is identified by contouring the projected stellar density (see Section 2.3 for details). The contrast ratio is defined as the the $95\%$ and $5\%$ percentiles of the stellar mass density within each clump contour. No-RP clumps (blue) tend to have high contrast ratios, whereas RP simulations (red) tend to have smaller contrast ratios.}
\label{fig:crmc}
\end{figure*}

The left column of Figure \ref{fig:crmc} shows the contrast ratio versus the clump mass for every clump in our sample for both RP (blue) and no-RP (red) simulations.  The contrast ratio is defined as the 95\% percentile highest density pixel divided by the 5\% percentile highest pixel enclosed by the clump boundary.  The middle column plots the contrast ratio, but against the clump mass normalised to the mass of the disk. Similarly, the right column plots the contrast ratio against the clump mass normalised to the total stellar mass of the galaxy. Errorbars denoting the 25\%, 50\% and 75\% limits in each mass bin are shown. 

The contrast ratio can be thought of as the peak density divided by the circumferential density which presumably traces the density nearby the clump. Because the contrast ratio can be non-parametrically measured on the set of projected pixels constituting a clump, similar measurements can be conducted observationally without the need to define a disk. Furthermore, choosing the 5\% percentile of the projected stellar mass density instead of the minimum is robust to small variations in the clump contouring.  Similarly, choosing the 95\% percentile instead of the maximum reduces susceptibility to spurious noise dominating the contrast ratio. Higher values of the contrast ratio indicate a more highly peaked density distribution, whereas lower values indicate a clump marginally distinguishable from the local disk density. Because the stellar clump finding method is intrinsically finding overdensities in the projected stellar mass maps, the contrast ratio is by construction $\geq 1.0$.  

For all of the mass definitions shown in Figure \ref{fig:crmc},  the median no-RP clumps have higher contrast ratios than the median RP clumps across all mass ranges. At low masses, the clumps marginally rise above the minimum detection threshold density, and so the contrast ratio is marginally above unity. For both RP and no-RP simulations, the contrast ratio for clumps climbs steadily as a function of clump mass regardless of normalisation. The median contrast ratio in no-RP simulations is $\sim1$ at $M_{clump}=10^7M_{\odot}$ and rises quickly to $\sim7$ at $M_{clump}=10^8M_{\odot}$ and reaches its maximum value of $\sim9$ at $M_{clump}=10^9M_{\odot}$.  Plots of the no-RP contrast ratio against the normalised clump masses $M_{clump}/M_{disk}$ and $M_{clump}/M_{total}$ mirror the trends in the raw clump mass, $M_{clump}$.  The contrast ratio for RP clumps also increases with clump mass, but much less rapidly than in no-RP simulations. At $M_{clump}=10^7M_{\odot}$ the median RP contrast ratio is 1.7,  and increases marginally to 2.0 at $M_{clump}=10^8M_{\odot}$. Between $M_{clump}=10^8M_{\odot}$ and $M_{clump}=10^9M_{\odot}$ the median contrast ratio climbs more rapidly and ultimately reaches a contrast ratio comparable with the no-RP contrast ratio. Thus, the median contrast ratio in no-RP simulations is systematically higher than RP simulations across all clump mass ranges, but converges for the highest mass clumps. 

\section[]{Clumps in Mock Observations}
\label{sec:mock}
In this section we extend the applicability of our findings to potentially observable clump statistics. We simultaneously study and find clumps in gas, stellar, and light maps. We qualitatively find that a minority of clumps are common to all three maps. We briefly discuss the salient properties of each clump finder, but do not attempt a rigorous and quantitative study of the differences between each clump finder. Furthermore, we find clumps in simulated mock observations and report that the number of clumps found in RP simulations is marginally lower than the number of clumps found in the no-RP simulations. We do not argue that the morphological effect of RP is sufficiently dramatic as to be bound by observational constraints.

\begin{figure*}
\includegraphics[scale=0.33]{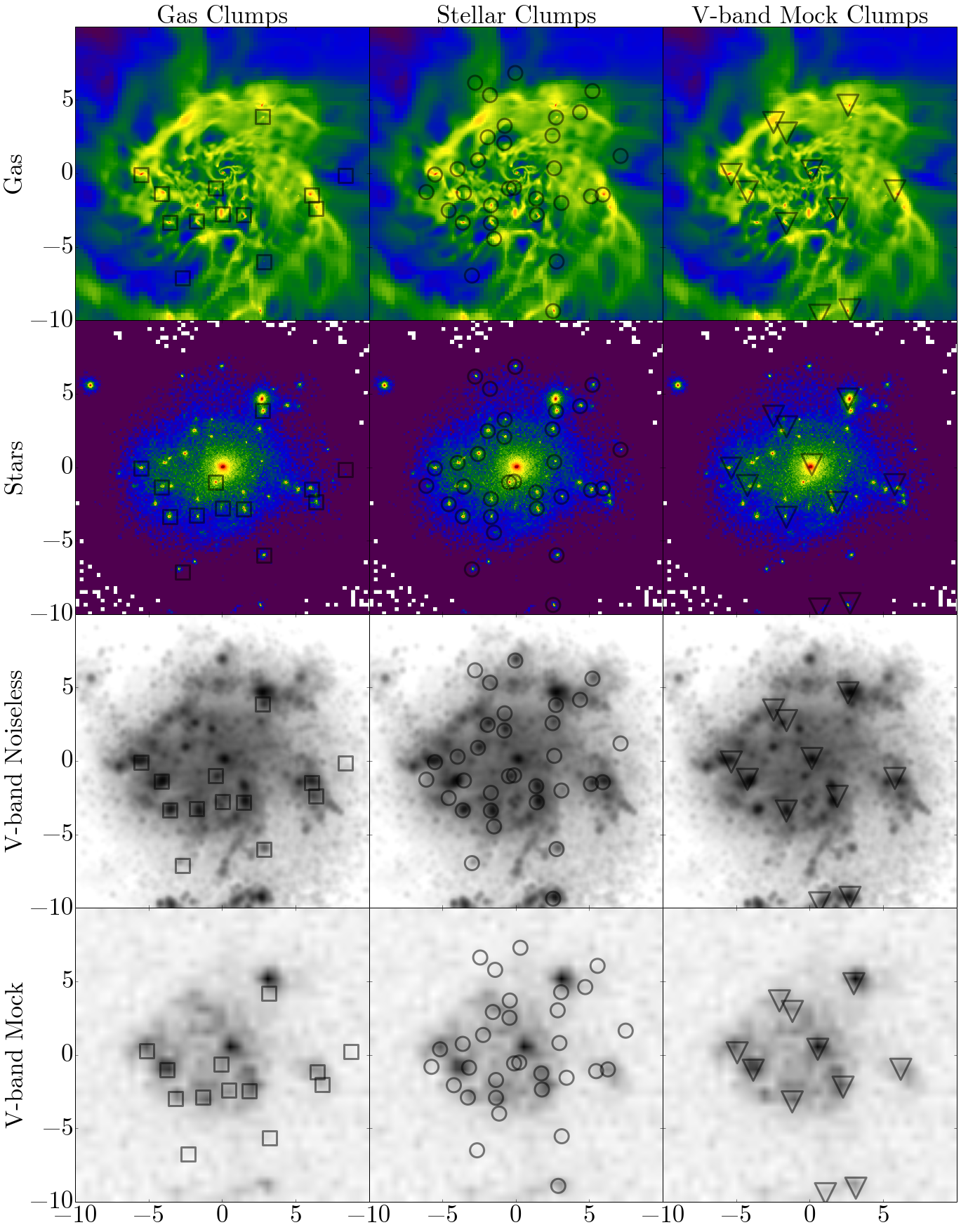}
\caption{Three clump finders, operating on three different projected quantities, are compared. The gas maps (first row), stellar mass maps (second row), simulated V-band images (third row), and mock V-band images (fourth row) are shown above. The third row shows the simulated V-band image before degrading. The bottom row images are degraded with background noise and an appropriate point-spread function. The mock images are comparable to HST restframe V-band observations. Clumps found in the gas (squares, first column from the left), stars (circles, second column), and V-band mock images (downward triangles, third column) are shown. Each map is $20$ kpc on a side. This mosaic demonstrates that clumps found in any single map are not necessarily clumps found in any other map although there are clumps common to multiple maps. Note that for gas and stellar maps, but not simulated V-band  and mock V-band images, clumps identified in the dark matter as minor mergers are excluded from the clump finder. As a result, the ongoing minor merger (top-right in every frame) is found but excluded from analysis in the gas and stellar clump finders, but not in the mock V-band clump finder.}
\label{fig:ecf}
\end{figure*}

In Figure \ref{fig:ecf} a matrix of four projected maps and three clump finders operating on the gas, stellar, and mock V-band maps is demonstrated. We only present this as an example of the clump finder performance and do not attempt a systematic comparison. The three clump finders automatically identify clumps that are also visually identified in the gas, stars, and light. Four maps of the VELA13 simulation at $a=0.330$ ($z=2.0$) are used in this comparison: the projected gas mass (top row), the stellar mass map (second row), the  simulated V-band images (third row), and finally the mock V-band images (fourth row). The simulated V-band images are the product of radiative transfer calculation via Sunrise, which includes sophisticated modelling of the stellar emission spectra and the absorption and scattering of light due to intervening dust. All V-band images are in the observed frame. The images in the bottom row are comparable with HST V-band observations since they have similar resolution and noise properties.

The first row shows the projected gas mass with clumps found using the algorithm discussed in \cite{2013arXiv1311.0013M}. Unlike the rest of the algorithms discussed here, this clump finder detects gas clumps in three dimensional space instead of the projected plane. The three dimensional region is smoothed at small scales comparable to the simulation resolution to remove irrelevant fluctuations and transient structures. The gas density field is also smoothed on the scale of the disk, thereby washing away small features, and the clump finding is computed on the residual density. This process of subtracting the estimated background promotes the contrast and identification of clumps. The centres of clumps found in the projected stellar mass are shown in the second column. The clump finding methods for the stellar mass maps and the V-band mock images are discussed in section \ref{clumpfinder}. The clump centres returned by the clump finder operating on the V-band mock images are shown in the third column. 

The centres of the clumps identified in this gas are shown as squares in the top-left diagram in Figure \ref{fig:ecf}. These clumps largely correspond to visually-identified gas clumps with a few spurious clumps detected (left column, first row). Furthermore, dark matter-dominated clumps are tagged and excluded as ex situ (e.g. minor mergers). The same clump locations are plotted on top of the stellar mass maps (left column, second row) as well as the simulated V-band (left column, third row) and mock V-band images (left column, fourth row). Out of the thirteen clumps found in this map, ten are also found in the stellar mass clumps. Nine out of these thirteen clumps are visually identified as also being clumps  in the simulated V-band image. However, in the mock V-band image, only one of the clumps is accurately recovered. Instead, the effect of smoothing and a PSF effectively blurs multiple nearby clumps and reduces the visibility of  isolated clumps. As a result, there is a poor correlation between individual clumps well-identified in the gas and those found in the V-band images.

The clumps found in the stellar mass are shown as circles in the second column of Figure \ref{fig:ecf}. These stellar clumps correspond well to visually-identified stellar clumps. However, out of the 34 clumps identified in the stellar mass maps, a majority have no analogous clumps in the gas maps. Most of the largest clumps are common to the gas map, but smaller clumps are less likely to be found in the gas map. 11 out of the 34 stellar mass clumps are not visually identified in the simulated V-band image. As with the gas clump finder, the low stellar mass clumps are less likely to be correlate with clumps found in other maps. Finally, none of the clumps identified in the stars exactly match clumps found in the mock V-band images. 

To find clumps in the mock images we use forthcoming methods currently being applied to HST observations as part of the CANDELS program (Guo et al., in prep.). A comparison with `by-eye' human visual classifications of clumpiness of the same observations (Mozena et al., in prep.) provides a sanity check for clump finding techniques. Generally speaking, both automated and human classifiers show good agreement, with the automated clump finder working effectively at detecting clumps.

We briefly detail the automated clump finding process in the mock images. In order to find clumps in the mock images, we first smooth a given mock image through a box car filter with a size of 10 pixels to obtain a smoothed image. Then, we subtract the smoothed image from the original image to make a contrast image. After measuring the background fluctuation from the contrast image with $\sigma$-clipping, we mask out all pixels below 2$\sigma$ of the background fluctuation to make a filtered image, where clumps stand out in a zero background. We then run SExtractor \citep{1996A&AS..117..393B} on the filtered image to detect sources, and exclude suspicious detections by enforcing a minimal detection area of 5 pixels. Each detected source is considered a clump. For more details on this method, please refer to Guo et al., in prep.

Clumps found in the mock V-band are shown in the third column as downward-facing triangles. As these are mock-observed images we do not associate clumps with a dark matter halo, and do not remove ex-situ clumps from the analysis as we have done with the gas- and star-based clump finders. There are three large clumps visible in the mock V-band image. The first clump, at $(x,y)=(3, 6)$, is an ongoing minor merger which is systematically excluded by the gas and stellar clump finders, but otherwise would have been found. The second clump, at $(x,y)=(1, 1)$, is the galaxy centre, which is similarly found and excluded by the gas and stellar clump finders. The third brightest clump, at $(x,y)=(-4, -1)$, is an object spanning a single clump identified in the gas, but is a separately identified clump in the stars. The `blurring' of the clumps is due to the combination of the smearing effect of the PSF, the addition of a noise background, and possibly dust obscuration of structural details. For example, the clump  at $(x,y) = (6, -1)$ is a combination of three clumps, two identified in gas and two identified in the stars. The net result is that none of the clumps found in the mock observations translate directly into a single clump observed in the raw simulation data.  

\begin{figure}
\includegraphics[scale=1.00]{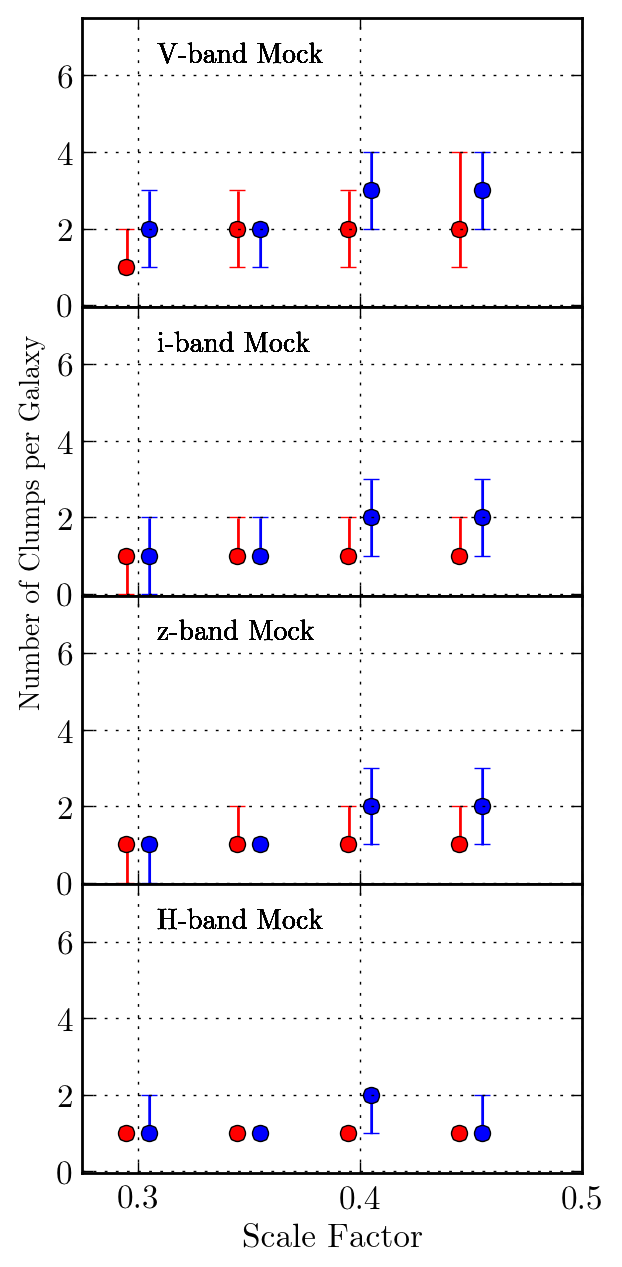}
\caption{The median number of clumps per galaxy is shown as a function of scale factor for a number of mock observed Hubble Space Telescope filters. In these cases, the image has been degraded with the appropriate point spread function (PSF) and background noise added. Also shown are errorbars representing the 25\% and 75\% percentiles. The filters are arranged from shortest wavelength (V band) to longest wavelength (H band). At early times in the images few clumps are found as the PSF and noise background conspire to blur and reduce the number of visible clumps. At later times the number of clumps increases, especially in the V band.}
\label{fig:guo}
\end{figure}

\begin{table}
    \begin{tabular}{llll}
    \toprule
    ~     & $f_{clumpy}$ & $f_{clumpy} $ & $f_{clumpy} $ \\ 
    ~     & $z=3.0-2.3$ &  $z=2.3-1.9$ & $z=1.9-1.5$ \\
    \midrule
    No-RP & 0.32               & 0.53                 & 0.64                 \\
    RP    & 0.24               & 0.48                 & 0.54                 \\
    \end{tabular}
    \caption{The fraction of  clumpy galaxies, $f_{clumpy}$ for no-RP and RP simulations in several redshift ranges. The clumpy fraction is the fraction of all simulations with one or more clumps counted in the mock V-band observations. RP simulations are overall less clumpy than no-RP simulations. }
\label{table2}

\end{table}

In Figure \ref{fig:guo} the median number of clumps in the mock V-band images is shown for RP and no-RP simulation as a function of scale factor.  The V-band observations are generally more sensitive to patchy star formation and clumps compared to the H-band observations, and thus their median number of clumps is higher. The median number of clumps is typically one at early times for most bands except the V-band. 
In Table \ref{table2} the clumpy fraction is given for RP and no-RP simulations for a range of redshifts. This fraction is defined as the number of V-band mock images with one or more off-center clumps divided by the total number of V-band mock images.
The V-band clump number, for both RP and no-RP is typically two or greater at all times. At late times the discrepancy between RP and no-RP grows, with the number of clumps in the no-RP being slightly greater.  As a result, we conclude that the number of observed clumps is affected by the inclusion of RP, albeit less dramatically than the effect of RP computed from raw simulated quantities. 

\section{Conclusions}

We study the effect of radiation pressure (RP) feedback on the total stellar mass and the formation of clumps in a suite of cosmological high-resolution zoom-in galaxy simulations. RP feedback suppresses the stellar mass of galaxies, particularly at early times, by a factor of $\sim2$ to better match constraints. Despite this reduction, median stellar masses for halos more massive than $5 \times 10^{11} M_{\odot}h^{-1}$ are still overproduced by a factor of $\sim3$ ($\sim1.5$) at z=2 (z=1.5) when compared to the abundance matching estimates of  \citet{2013ApJ...770...57B}.

We find that simulations with RP depress the average number of low-mass clumps dramatically when compared to analogous simulations without RP. From $z=2.3$ to $z=1.9$ the average number of clumps in no-RP simulations is $\sim6$ but rises to $\sim9$ from $z=1.5$ to $z=1.0$. The inclusion of RP does not affect all clumps, but for RP simulations the stellar mass clump finder finds only 1.7 clumps on average, rising to 1.8 clumps over the same redshift intervals. However, these average clump counts include many low mass clumps, and above a threshold of $M_{clump}/M_{disk} > 5\%$ clump counts in the RP and no-RP simulation are comparable. That the counts are comparable above this limit suggests that  RP, while affecting smaller clumps significantly, does not similarly impact large clumps. At these high masses, the ratio of RP to no-RP clumps rises to unity, implying that the effect of radiation pressure on massive clump statistics is negligible. The contrast ratios of no-RP clumps are higher than RP clumps by a factor of a few for low mass clumps, but the differences in contrast ratio diminish with increasing clump mass. In general we find that the differences between clump properties in RP and no-RP simulations converge for these high mass clumps.  We note however, that the simulations analysed are relatively low-mass galaxies with disk stellar masses below about $2 \times 10^{10} M_\odot h^{-1}$. 

By processing these simulations through the Sunrise radiative transfer code, we create mock observations analogous to CANDELS observations and characterise the effect of RP on mock observed clumps. In all wavebands, at all redshifts, in the mock images the number of clumps in the no-RP case exceeds that of the RP case. However, the observed difference in clump counts, while stark when viewed in the projected stellar mass, is negligible in the mock observations. While RP does reduce the number of clumps in the observations, particularly at $z\sim1$ and in the V-band, the reduction is small and not in disagreement with observations of clumpy galaxies. We also qualitatively report that a poor correlation exists between clumps found in the stellar mass maps, gas maps, and mock V-band maps. 

\section*{Acknowledgments}
The simulations were conducted at the NASA Advanced Supercomputing (NAS) at NASA Ames Research Center and at the National Energy Research Scientific Computing Center (NERSC) at Lawrence Berkley National Laboratory. We would like to acknowledge fruitful conversations with Sandra Faber and Matthew Turk.  YG and DCK acknowledge support from NSF AST-0808133. CEM, AD, and JP acknowledge support for grants NSF-AST-1010033 and HST 12060.12-A. DC is a Juan-de-la-Cierva fellow supported by AYA2012-31101 grant. 

\bibliographystyle{mn2e}
\bibliography{bib}

\end{document}